\documentstyle[prl,aps,psfig,epsf]{revtex}
\begin{document}
\draft
\title{Phase-coherent effects in multiterminal superconductor/normal metal
mesoscopic structures}
\author{A.F. Volkov$^{\ast \dagger }$ and R. Seviour$^{\ast }$,}
\address{$^*$ School of Physics and Chemistry,\\
Lancaster University, Lancaster LA1 4YB, U.K.\\
$^{\dagger}$Institute of Radioengineering and Electronics of the Russian \\
Academy of Sciencies, Mokhovaya str.11, Moscow 103907, Russia.}
\date{\today}
\maketitle

\begin{abstract}
In this report we analyse three effects which may arise in a mesoscopic
multiterminal S/N structure (two normal and two superconducting reservoirs).
We show that the Josephson critical current is a non-monotonic function of a
voltage between the normal reservoirs. The influence of the spin-polarized
electrons on the Josephson critical current was also studied. We show that
if there is a temperature difference between the normal reservoirs, a
voltage between the superconductors and  normal conductors arises which
oscillates as the phase difference varies. Its magnitude is much greater
than the thermoemf in the case of the ordinary thermoelectric effect.
\end{abstract}

\bigskip {\bf 1. Introduction}

In this report we present results of the theoretical study of transport in
diffusive 4-terminal superconductor-normal metal (S/N) mesoscopic
structures. A great deal of attention was paid recently to the problem of
the influence a current between the N reservoirs has on the Josephson effect
in the superconducting curcuit. This problem was analysed first in Ref.\cite
{Bart1} for a ballistic S/N system and in Ref. \cite{Vprl1} for a diffusive
system. It was shown that the Josephson critical current $I_{c}$ changes
sign at a certain voltage $V_{\text{ \ }}$between the N reservoirs, and the
Josephson junction switches into a $\pi $-state ($\pi $-junction). The
reason for the sign reversal effect is that the distribution function in the
N wire, which can have an equilibrium form, is shifted with respect to the
distribution function in the superconductors by the value $V$. Later the
dependence $I_{c}(V)$ was calculated in various models and approximations in
Refs. \cite{VT,Vmicro,Yip,W2,Wmicro} (diffusive case) and \cite{Shum,Bag}
(ballistic case). The sign reversal effect in a diffusive Au/Nb mesoscopic
structure was observed by Baselmans et al. \cite{Bart2}.

In multi-terminal S/N structures one can observe not only the sign reversal
effect, but also a number of other interesting phenomena. It was shown in
Refs. \cite{VT} that the measured critical current $I_{m}$ in a structure
similar to that shown in Fig.1a may not only decrease, but also may increase
with increasing voltage $V$. In particular one can observe the
Josephson-like effects (plateau on the $I_{3}(V_{S})$ curve, oscillations of
the measured critical current $I_{m}$ in a magnetic field etc) even if the
Josephson coupling between superconductors under equilibrium conditions is
vanishing. The stimulation of the Josephson effect in this structure is
related in particular to a phase dependence of the S/N interface or N wire
conductance: $\delta G\sim \cos \varphi .$ It turns out that apart from the
Josephson term $I_{c}\sin \varphi ,$ an additional term $I_{sg}\cos \varphi $
appears in a dynamic equation for $\varphi $. This term leads to a change of
the critical current from the value $I_{c}$ to a new (measured) value $I_{m}=%
\sqrt{I_{c}^{2}+I_{sg}^{2}}$,where $I_{sg}$ is the amplitude of the subgap
current. The stimulation of the Josephson effect by a current between the N
reservoirs was analysed in Refs.\cite{VT} on a simplified model of gapless
superconductors. Here we present results of the analysis for the real case
where the superconductors have a gap and the density-of-states (DOS) has a
singularity \cite{SV}. We show that the measured critical current $I_{m}$
reaches a maximum at $eV \cong \Delta $ and depends weakly on the
temperature $T.$ These results agree with recent observations of the
stimulation of the Josephson critical current \cite{Denm}.

We also study the effect of the spin injection into the N wire from the
ferromagnetic (F) reservoirs on the critical current $I_{c}$ for the
structure shown in Fig.1b. If the magnetizations in the F reservoirs have
the same direction, then there is no spin injection. In the case of
antiparallel orientation, the total magnetization of the injected electrons
in the N wire is finite. We show that in the case of a N wire short length $%
2L$ ($L<\{L_{\epsilon },L_{sp}\}$; here $L_{\epsilon }$ and $L_{sp}$ are the
energy and spin relaxation length) the spin injection does not affect the
critical current and the dependence $I_{c}(V)$ is the same as in the case of
normal (nonmagnetic) reservoirs. In a semi-mesoscopic structure $%
(L_{\epsilon }<L<L_{sp})$ the $\pi $-junction is realised only in the case
of the antiparallel orientation of the magnetizations in the F reservoirs
when spins are accumulated in the N wire.

Finally we analyse the thermoelectric effect in the structure shown in
Fig.1a. We assume that there is no current between the N reservoirs, but the
reservoirs are maintained at different temperatures $T=T_{o}\pm \delta T$.
It will be shown that the temperature gradient leads to a voltage difference 
$V_{T\text{ }}$between the normal and superconducting circuits. This voltage
does not contain the small parameter $T/\epsilon _{F}$ (as is the case for
the ordinary thermoeffect in normal metals) and reaches a value of order $%
\delta T/e$. In addition, the voltage $V_{T\text{ }}$ depends on the phase
difference $\varphi $; it is zero at $\varphi =0$ and oscillates with
increasing $\varphi .$

{\bf 2. Basic equations}

The Green's function technique is perhaps the most convenient and powerful
method for studing transport and nonequlibrium phenomena in S/N mesoscopic
structures. If we are not interested in purely quantum interference effects
arising as the result of multiple (non-Andreev) reflections, then we can use
a simpler, well developed method of quasiclassical\ Green's functions (see,
for example, \cite{LO}). In this method the matrix Keldysh function $%
\widehat{G}$ is introduced along with the retarded (advanced) matrix Green's
functions $\widehat{G}^{R(A)}=G^{R(A)}\widehat{\tau }_{z}+$ $\widehat{F}%
^{R(A)}$, here $\widehat{F}^{R(A)}$ is the matrix condensate Green's
function. The functions $\widehat{G}^{R(A)}$determine the excitation
spectrum and the DOS. The function $\widehat{G}$ describes the
nonequilibrium properties and is related to the matrix distribution
functions $\widehat{f\text{ }}$: $\widehat{G}_{\alpha }=\widehat{G}_{\alpha
}^{R}\widehat{f\text{ }}_{\alpha }-\widehat{f\text{ }}_{\alpha }\widehat{G}%
_{\alpha }^{A}$, where $\widehat{f\text{ }}_{\alpha }=f_{\alpha +}\widehat{1}%
+f_{\alpha -}\widehat{\tau }_{z}.$ The components of $\widehat{f\text{ }}%
_{\alpha }$ can be expressed through the distribution functions of electrons 
$n_{\alpha }$ and holes $p_{\alpha }$ ($\alpha $ is the spin index). The
function $f_{\alpha +}$ describes an electron and hole distribution
symmetrical in branch populations: $f_{\alpha +}=1-(n_{\alpha }+p_{\overline{%
\alpha }})$, here $\overline{\alpha }=\downarrow $ if $\alpha =\uparrow $.
The function $f_{\alpha -}=-(n_{\alpha }-p_{\overline{\alpha }})$ describes
the branch imbalance \ and determines the electric potential.

The equations for $\widehat{G}_{\alpha }$ and $\widehat{G}_{\alpha }^{R(A)}$
can be solved analitically in the case of a short structure ($\epsilon
_{Th}\equiv D/L^{2}>>\epsilon $, where $\epsilon $ is a characteristic
energy; $\epsilon \cong \min \{T,\Delta \}$)\cite{AVZ1,AVZ2,Z2,VZKl,Naz3}.
In the general case one needs to solve equations for the distribution
functions $f_{\alpha \pm }$ and the Usadel equation \cite{VZKl,AL,AsV,SmS}.
In the present paper we consider the case when the temperature may be both
greater or less than the Thouless energy $\epsilon _{Th}$ and the
approximation of a short length is not valid$.$ The distrubution functions
in the structure shown in Fig.1a obey the kinetic equation (we consider only
the dirty limit)

\begin{equation}
L{\Large \partial }_{x}[{\Large M}_{\pm }{\Large \partial }_{x}f_{\pm }(x)%
{\Large +J}_{S}f_{\mp }(x)\pm {\Large J}_{an}{\Large \partial }_{x}f_{\mp
}(x)]{\Large =r[}\stackrel{}{A_{\pm }}{\Large \delta (x-L}_{1}{\Large )+}%
\stackrel{\_}{A_{\pm }}{\Large \delta (x+L}_{1}{\Large )].}
\end{equation}
where $\ M_{\pm }=(1-G^{R}G^{A}\mp (\widehat{F}^{R}\widehat{F}^{A})_{1})/2;$ 
$J_{an}=(\widehat{F}^{R}\widehat{F}^{A})_{z}/2,\ J_{s}=(1/2)(\widehat{F}%
^{R}\partial _{x}\widehat{F}^{R}-\widehat{F}^{A}\partial _{x}\widehat{F}%
^{A})_{z},$\ $A_{\pm }=(\nu \nu _{S}+g_{1\mp })(f_{\pm }-f_{S\pm })-(g_{z\pm
}f_{S\mp }+g_{z\mp }f_{\mp });g_{1\pm }=(1/4)[(\widehat{F}^{R}\pm \widehat{F}%
^{A})(\widehat{F}_{S}^{R}\pm \widehat{F}_{S}^{A})]_{1};g_{z\pm }=(1/4)[(%
\widehat{F}^{R}\mp \widehat{F}^{A})(\widehat{F}_{S}^{R}\pm \widehat{F}%
_{S}^{A})]_{z};$

The parameter $r=R/R_{b}$ is the ratio of the resistance of the N wire $R$
and S/N interface resistance $R_{b}$; the functions $\stackrel{\_}{A_{-}}$
and $\stackrel{\_}{A}_{+}$ coincide with $\stackrel{\_}{A_{-}},\stackrel{\_}{%
A}_{+}$ if we make a substitution $\varphi \rightarrow -\varphi $. We
introduced above the following notations $(\widehat{F}^{R}\widehat{F}%
^{A})_{1}=Tr(\widehat{F}^{R}\widehat{F}^{A}),$ $(\widehat{F}^{R}\widehat{F}%
^{A})_{z}=Tr(\widehat{\tau }_{z}\widehat{F}^{R}\widehat{F}^{A})$ etc.; $\nu
, $ $\nu _{S}$ are the DOS in the N film at $x=L_{1}$ and in the
superconductors. The functions $f_{S\pm }$ are the distribution functions in
the superconductors which are assumed to have equilibrium forms. This means
that $f_{S+}\equiv f_{eq}=\tanh (\epsilon \beta )$ (here $\beta =1/(2T)$)
and $f_{S-}=0$, because we set the potential of the superconductors equal to
zero (no branch imbalance in the superconductors). We assumed that the width
of the S/N interface $w$ is much less than $L_{1,2}$ and introduced the $%
\delta -$functions. Note that in some papers the last term in the left-hand
side of Eq.(1) is missing. One can solve Eq.(1) and express the distribution
functions $f_{\alpha \pm }$ in terms of the retarded (advanced) Green's
functions which should be found from the Usadel equation. Integrating Eq.(1)
once, we obtain

\begin{equation}
{\Large M}_{\pm }{\Large \partial }_{x}f_{\pm }(x){\Large +J}_{S}f_{\mp
}(x)\pm {\Large J}_{an}{\Large \partial }_{x}f_{\mp }(x)={\Large J}_{1,2\pm }
\end{equation}
where the indeces $1,2$ relates to the intervals $(0,L_{1})$,$(L_{1},L)$ and
the constants $J_{1,2\pm }$ are partial flows. For example, the partial flow 
$J_{1,2-}$ determines the electrical current in the intervals $(0,L_{1})$,$%
(L_{1},L)$

\begin{equation}
I_{1,2}=\sigma \int_{0}^{\infty }d\epsilon J_{1,2-}
\end{equation}
Here $\sigma $ is the conductivity of the N wire. As it follows from Eq.(1),
the flows $J_{1,2\pm }$ are connected with each other

\begin{equation}
{\Large J}_{2\pm }-{\Large J}_{1\pm }=rA_{\pm }
\end{equation}

The functions $rA_{\pm }$ are the flows through the S/N interface. In
particular, $rA_{-}$ is the partial electrical current through the S/N
interface. The first term $(\nu \nu _{S}+g_{1+})f_{-}$ in the expression for 
$A_{-}$ is the quasiparticle current (the term $g_{1+}f_{-}$ contributes to
the subgap current), and the second term $(g_{z+}f_{S+}+g_{z-}f_{+})$ is
related to the supercurrent. One can show that the partial supercurrent $%
J_{S}$ is a constant in the interval $(0,L_{1})$ and equals zero outside
this interval.

We can solve Eqs.(2) and express the distribution functions $f_{\pm }$
through the constants $J_{1,2\pm }.$ At $x=\pm L$ the functions $f_{\pm }$
are equal to equilibrium distribution functions in the N reservoirs.
Therefore we can find the functions $f_{\pm }(x)$ and substituting them into
Eq.(4) determine $f_{\pm }(L_{1})$ and $J_{1,2\pm }$. Thus the problem is
reduced to solving the Usadel equation. The solution of the Usadel equation
can be found either numerically or analitically in limiting cases. In the
next sections we will employ this method to study the dependence of the
critical current on the voltage between the N reservoirs and the
thermoeffect.

{\bf 3. Suppression and enhancement of the critical current}

In this section we consider a symmetrical structure shown in Fig.1a. We
assume that the electric potential at the N reservoirs is $V$ and at the
superconductors is zero (the quasiparticle currents flow from the N
reservoirs to the superconductors and the supercurrent flows between the
superconductors). The distribution functions in the reservoirs have the
equilibrium form: $F_{V\pm }=[\tanh ((\epsilon +eV)\beta )\pm \tanh
((\epsilon -eV)\beta )]/2.$

First we consider the case of a small $r,$ that is, the interface resistance
is large and the condition

\begin{equation}
(\epsilon \tau _{\epsilon })^{-1}<<r<<1
\end{equation}
should be satisfied. The first condition ensures that inelastic collision
term can be neglected (here $\tau _{\epsilon }$ is the energy relaxation
time, $\epsilon \approx \min \{T,\epsilon _{Th}\}$). Then the distribution
function $f_{+}$ is constant and in the main approximation is $f_{+}\cong
(F_{V+}+f_{eq}(r_{2}\nu \nu _{s}))/(1+r_{2}\nu \nu _{s}),$ here $%
r_{2}=rL_{2}/L$. The function $f_{-}$ is constant in the interval $(0,L_{1})$
and in the main approximation equals $f_{-}\cong F_{V-}/(1+r_{2}\nu \nu
_{s}) $. Outside this interval the function $f_{-}$ increases linearly to
the value $F_{V-}.$ Following the method presented above, we find the
current through the S/N interface

\begin{equation}
I_{3}(V)=I_{2}-I_{1}=I_{o}(V)+I_{sg}(V)\cos \varphi -I_{c}(V)\sin \varphi
\end{equation}

The first two terms are the quasiparticle current and the last term is the
Josephson supercurrent. All the components of the current depend on $V$.
This expression shows that at a given control voltage $V$ and zero voltage
difference between superconductors ($\varphi $ is constant in time) the
current $I_{3}$ may vary with changing $\varphi $ in some limits:$\mid
I_{3}(V)-I_{o}(V)\mid \leq I_{m}(V)$. This means a plateau on the $%
V_{S}(I_{3})$ characteristics (see \cite{VT}); here $V_{S}=(\hbar
/2e)\partial _{t}\varphi $ is the voltage difference between
superconductors. We can write the phase-dependent part of $I_{3}$ in the
form $I_{3\varphi }=I_{m}\sin (\varphi +\alpha )$, where $I_{m}=\sqrt{%
I_{c}^{2}+I_{sg}^{2}}$ is the measured critical current, $\cos \alpha
=I_{c}/I_{m}$. In Fig.2 we plot the dependence of the measured critical
current $I_{m}$ on the control voltage $V$ for different temperatures. It is
seen that the temperature dependence of the maximum of $I_{m}$ which is
achieved at $eV\simeq \Delta $ is much weaker than the $I_{c}(T)$
dependence. These results qualitatively agree with the experimental data 
\cite{Denm}.

{\bf 4}. {\bf Spin injection and the critical Josephson current.}

In this section we present results of the study of spin injection on the
critical Josephson current $I_{c}$ for the structure shown in Fig.1b \cite
{VFa}. We calculate the distribution function $f_{+}$ which determines $%
I_{c} $ for parallel $(\uparrow \uparrow )$ and antiparallel $(\uparrow
\downarrow )$ orientations of the magnetization in the ferromagnetic (F)
reservoirs. It is assumed that the voltage $2V$ is applied between the F
reservoirs and a current flows between the N reservoirs. We consider
different limits of the length of the N wire.

{\it 4.1. Mesoscopic limit:} $L<\{L_{\epsilon },L_{sp}\}.$

In this limit the function $f_{+}\equiv (f_{\uparrow +}+f_{\downarrow +})/2$
does not depend on mutual orientation of ferromagnets and has the same form
as in the nonmagnetic case: $f_{+}=F_{V+}.$ The conductance $G$ and magnetic
moment of injected spins $M$ (per unit volume of the N wire) depend on
mutual orientation of the ferromagnetic domains. In the case of the parallel
orientation $(\uparrow \uparrow )$ we obtain $G=(2R_{b\uparrow
}+R_{L})^{-1}+(2R_{b\downarrow }+R_{L})^{-1}$ and $M=0$. In the case of the
antiparallel orientation $(\uparrow \downarrow )$ we have $G=2/(R_{b\uparrow
}+R_{b\downarrow }+R_{L})$ and $M=4\mu (R_{b\uparrow }-R_{b\downarrow
})V\beta /\sigma ^{2}(2R_{b\uparrow }+R_{L})(2R_{b\downarrow
}+R_{L})(R_{b\uparrow }+R_{2\downarrow }+R_{L}).$ Here $R_{b\uparrow
,\downarrow }$ and $R_{L}$ are the F/N interface and the N wire resistances
for electrons with up and down spins . Since the interface conductances $%
R_{b\uparrow ,\downarrow }^{-1}$ differ from each other (they are
proportional to different density of states in the ferromagnets $\nu
_{F\uparrow ,\downarrow }$), spin injection takes place in the antiparallel $%
(\uparrow \downarrow )$ configuration. However the critical current $I_{c}$
is the same for both configurations. Its dependence on $V$ is similar to
that found for the structure under consideration with normal (nonmagnetic)
reservoirs \cite{Yip}.

{\it 4.1. Semi-mesoscopic limit:} $L_{\epsilon }<L<L_{sp}.$

In this limit there is no spin relaxation in the N wire and the
magnetization $M$ is determined again by formulae presented above. On the
other hand the magnetization is related to a potential $V_{sp}$ which
determines an imbalance between the spin subsystems in the N wire: $M=2\mu
\nu eV_{sp}$. The function $f_{+}\equiv (f_{\uparrow +}+f_{\downarrow +})/2$
has an equilibrium form corresponding to the potential $V_{sp}:f_{+}=[\tanh
(\epsilon +eV_{sp})\beta +\tanh (\epsilon -eV_{sp})\beta ]/2$. Therefore in
the case of parallel $(\uparrow \uparrow )$ orientation $I_{c}$ does not
depend on $V$ as $V_{sp}=0$ ($V$ is proportional to $V_{sp}$)$.$ In the case
of antiparallel $(\uparrow \downarrow )$ orientation $I_{c}$ does depend on $%
V_{sp}$ in the same way as in the section 4.1. A similar effect of spin
injection into a supereconductor on the energy gap $\Delta $ was analysed in
Ref. \cite{Jap}. The critical Josephson current is more sensitive to the
form of the distribution function than the energy gap $\Delta $. Therefore
the effect considered here allows one to realise the $\pi -$junction and to
make certain conclusions about the energy and spin relaxation rate.

{\bf 5. Giant thermoelectric effect}

It is well known that if the terminals of a normal conductor are maintained
at different temperatures $\delta T=T_{o}\pm \delta T$, then in the absence
of the current an thermoemf $V_{emf}$ appears at the terminals. The
magnitude of $V_{emf}$ is equal to: $eV_{emf}=c_{1}(T/\epsilon _{F})\delta
T, $ where $c_{1}$ is a factor of the order 1 and $\epsilon _{F}$ is the
Fermi energy. In this report we study the thermoelectric effect in the
mesoscopic structure shown in Fig.1a. It will be shown that a temperature
difference between the normal (N) reservoirs leads to a voltage between
normal and superconducting curcuits $V_{T}$. The magnitude of this voltage
does not contain the small parameter $(T/\epsilon _{F}),$ and besides it
oscillates with a variation of the phase difference $\varphi $ between the
superconductors.

We assume that the superconductors are connected via a superconducting loop
and the phase difference between them $\varphi $ is controlled by an applied
magneitic field. The N reservoirs are disconnected and maintained at
different temperatures $T(\pm L)=T_{o}\pm \delta T$. We will calculate the
electric potential in the N film and in particular the potential $V_{T}$ in
the N reservoirs . Since we set the potential in the superconductors equal
to zero, the potential $V_{T}$ is the voltage difference between the N
reservoirs and superconductors which arises in the presence of the
temperature difference $\delta T.$ In the limit of high interface resistance
(see condition (5)) we obtain from Eq.(2) in the main approximation in $r:$ $%
f_{+}(x)\cong J_{+}x+f_{+}(0),$ where $f_{+}(0)=f_{eq}\equiv \tanh (\beta
_{o}\epsilon ),$ and $J_{+}=\delta \beta \epsilon /L\cosh ^{2}(\beta
_{o}\epsilon ),$ $\delta \beta =-\beta \delta T/T_{o}.$ We assumed the ratio 
$\delta T/T_{o}$ to be small and expant the distribution functions in the N
reservoirs.

In order to find the function $f_{-},$ we consider Eq.(2) and take into
account that $J_{2-}$ is zero (no current through the N reservoirs). In the
absence of the temperature gradient we obtain from this equation

\begin{equation}
f_{-}=0,\text{ \ }J_{S}f_{eq}=J_{1eq}=r(g_{z-}+g_{z+})f_{eq}
\end{equation}

The temperature gradient leads to the non-zero function $f_{-}(x)$ which
determines the electrical potential $V_{T}(x)$. We find from Eq.(2)

\begin{equation}
f_{-}(x)=f_{-}(0)+\delta J_{1}x+\int_{0}^{x}dx_{1}[{\Large J}_{an}{\Large %
\partial }_{x}\delta f_{+}(x_{1})-{\Large J}_{s}\delta f_{+}(x_{1})]
\end{equation}
where $\delta f_{+}(x)\equiv f_{+}(x)-f_{eq}$. From Eq.(1) we obtain for $%
\delta J_{1}\equiv J_{1}-J_{1eq}$

\begin{equation}
\delta J_{1}=\pm r[g_{z+}\delta f_{+}(\pm L_{1})-g_{1+}f_{-}(\pm L_{1})]
\end{equation}
Here the signs $\pm $ relate to the points $\pm L_{1}$. From Eqs.(8) and (9)
it follows that $\delta J_{1}=0,$ and in the main approximation in $r$ the
distribution function is $f_{-}(0)\approx \delta f_{+}(\pm
L_{1})g_{z+}/g_{1+},$ where $\delta f_{+}(\pm L_{1})=\pm J_{+}L_{1}$. The
electric potential $V_{T}$ is related to the function $f_{-}$ and is an even
function of $x$

\begin{equation}
eV_{T}=\int_{0}^{\infty }d\epsilon f_{-}=\frac{L_{1}}{L}\frac{\delta T}{T_{o}%
}\beta \int_{0}^{\infty }d\epsilon \frac{g_{z+}}{g_{1+}}\frac{\epsilon }{%
\cosh ^{2}(\beta _{o}\epsilon )}
\end{equation}

Here we substitute the function $f_{-}(0)$ which is the main contribution to
the function $f_{-}(x).$ Therefore the potential $V_{T}$ arising in the
presence of the temperature gradient is almost constant along the N wire.
This potential equals approximately the voltage difference $V_{T}$ between
the N reservoirs and superconducting loop. It is worth noting that $V_{T}$
determined by Eq.(10) does not depend on the small parameter $r$ (however
one should have in mind that the condition (5) imposes limits on this
parameter). We calculated the integrand using both analytical an numerical
solutions of the Usadel equation. In Fig.3 we show the temperature
dependence of $V_{T}$ for $\varphi =\pi /2$. We see that the dependence $%
V_{T}(T)$ is nonmonotonic (reentrant behaviour). One can easily estimate $%
V_{T}$ on the order of magnitude. We obtain

\begin{equation}
eV_{T}=\delta T(L_{1}/L)\sin \varphi \left\{ 
\begin{array}{c}
(T/\epsilon _{Th}){\it C}_{1}(\varphi ),\text{ \ \ }T<<\epsilon _{Th} \\ 
(\epsilon _{Th}/T)^{2}{\it C}_{2}(\varphi ),\text{ \ }T>>\epsilon _{Th}
\end{array} \right.
\end{equation}

Here ${\it C}_{1,2}(\varphi )$ are periodic functions of the phase
difference $\varphi $; they are of order 1 and are not zero at $\varphi =0.$
If the ratio $T/\epsilon _{Th}$ is of order 1, then the thermoemf is of the
order $\delta T$, that is, several orders greater than the thermoemf in the
normal metals.

The physical explanation of the effect is the following. The temperature
gradient creates a deviation of the distrubtion function $\delta
f_{+}=-\delta (n+p)$ from equilibrium. On the other hand the superconductors
do not affect this function because complete Andreev reflections conserve
the total number of excess electrons and holes. The function $\delta f_{+}$
has different signs at $\pm L_{1}$ and leads to an additional Josephson
current $\delta J_{S}=rg_{z+}\delta f_{+}(L_{1})$ of the same sign at $\pm
L_{1}$(sign of the function $g_{z+}$ is different at these points). In order
to compensate these additional currents, the potential $V_{T}$ arises in the
N wire producing a subgap current $\delta J_{sg}=rg_{1+}f_{-}(\pm L_{1})$
which compensates the current $\delta J_{S}$.

We have neglected the ordinary thermoelectric current $I_{T}$ in the N wire
because it contains a small parameter $(T/\epsilon _{T})$ and leads to a
small contribution to $V_{T}.$ Recently the influence of the proximity
effect on the ordinary thermoelectric effect in the mesoscopic S/N structure
was studied both theoretically \cite{Lam} and experimentally \cite{Chandr}.
The effect analysed here is much stronger.

\bigskip

In simplified models we have analysed three possible effects in the
4-terminals S/N mesoscopic structures; one of them (see section 3) has been
observed experimentally. Further theoretical studies of these and other
effects are needed to account for processes which were neglected in our
analysis. It is interesting, for example, to investigate how the obtained
results are changed by the energy relaxation in the N wire which seems to be
faster than it was expected \cite{Estev}

We are grateful to the EPSRC for their financial support.

\begin{figure}
\vspace{0.3cm}
\hspace{3.2cm}
\epsfxsize=6cm
\epsfysize=6cm
\epsffile{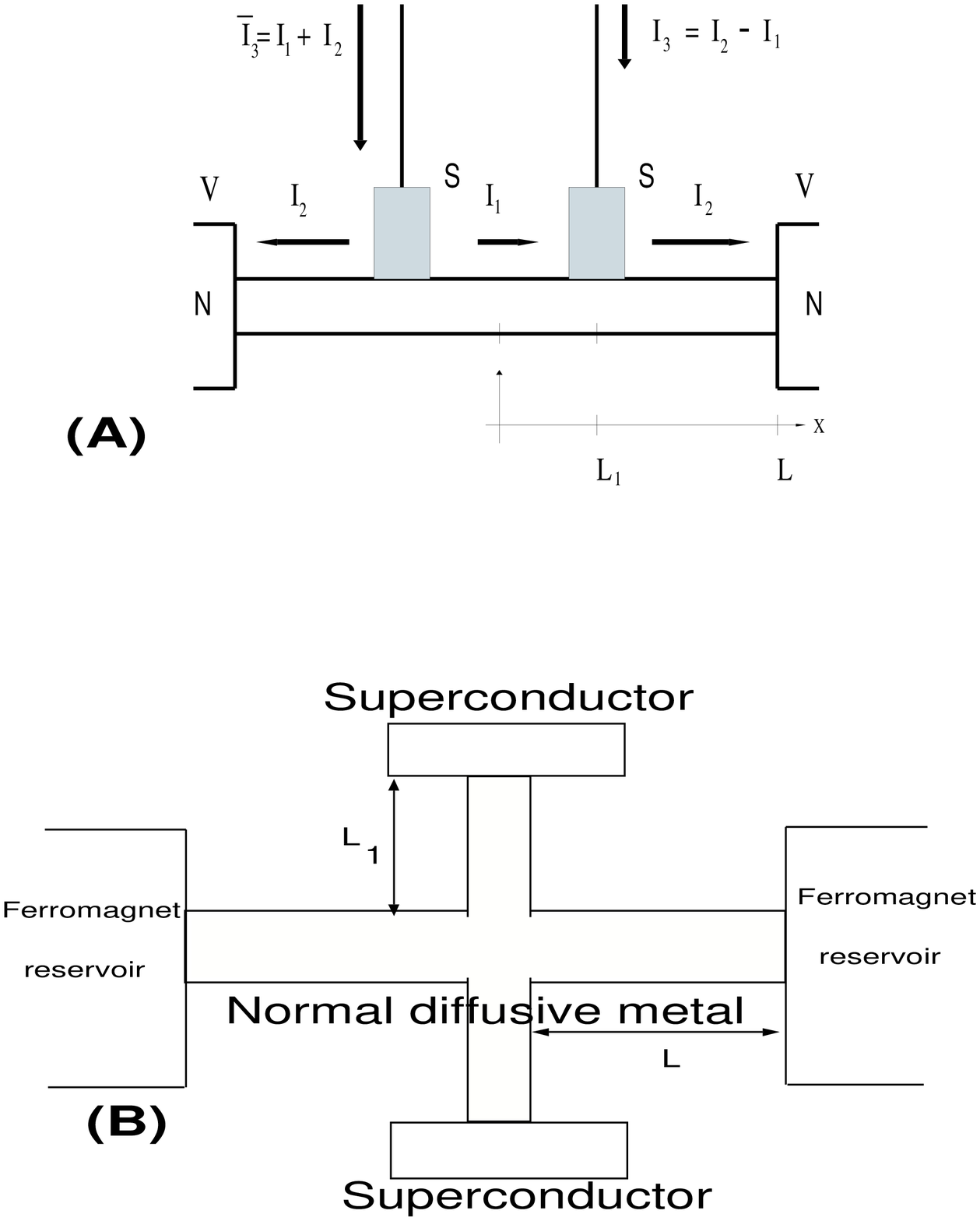}
\vspace{0.3cm}
\refstepcounter{figure}
\label{fig1}
\end{figure}
{\small \setlength{\baselineskip}{10pt} FIG.\ \ref{fig1} 
Schematic view of the 4-terminal S/N/S structure under consideration.}

\begin{figure}
\vspace{0.3cm}
\hspace{3.2cm}
\epsfxsize=6cm
\epsfysize=5cm
\epsffile{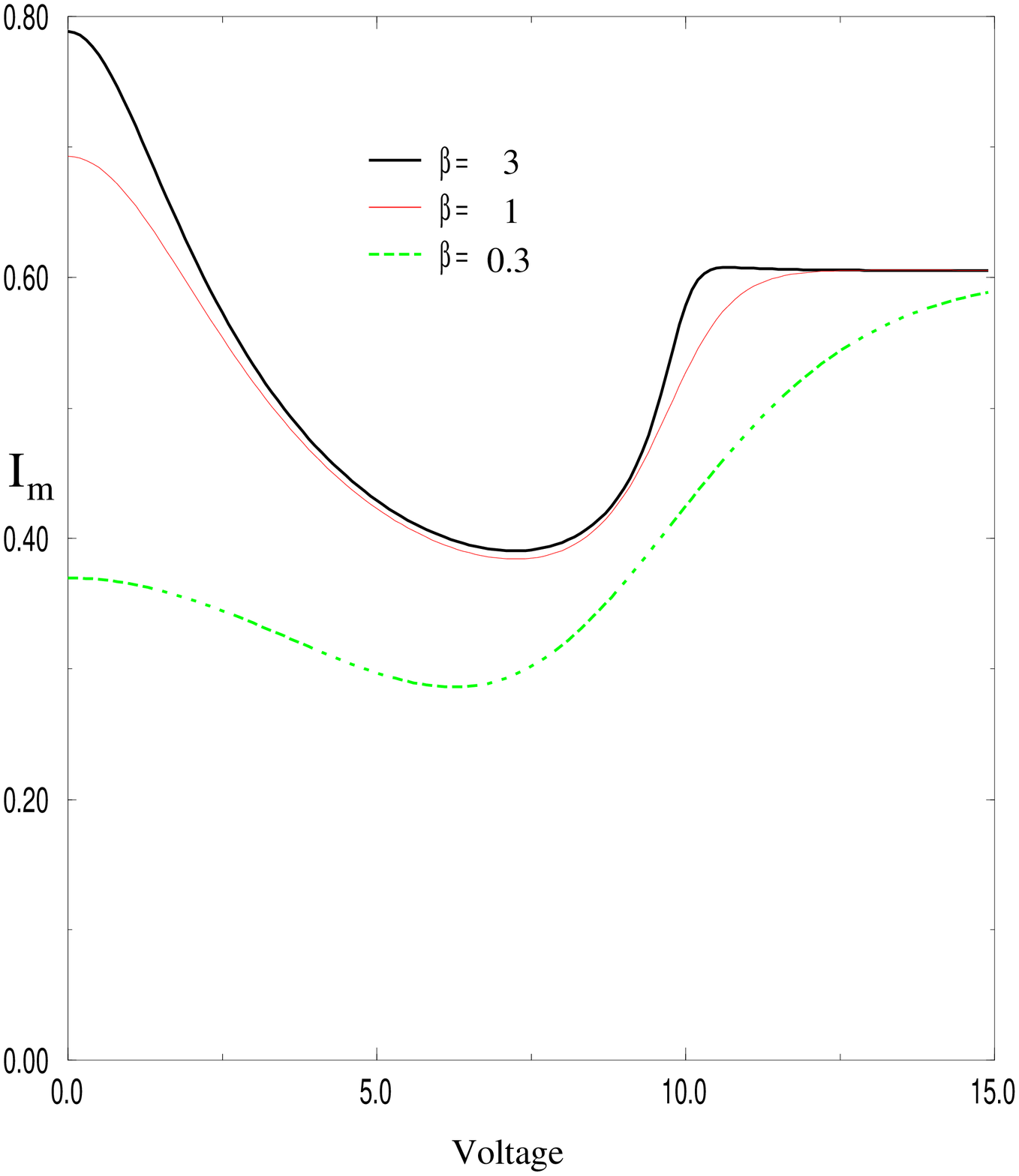}
\vspace{0.3cm}
\refstepcounter{figure}
\label{fig3}
\end{figure}
{\small \setlength{\baselineskip}{10pt} FIG.\ \ref{fig3} 
 The measured critical current ($I_{m}$) vs $V$ for different
temperatures: $\beta =\epsilon _{Th}/2T$. The parameters are: $\Delta
=10\epsilon _{Th},L_{1}/L=0.3, r=0.3$. }

\begin{figure}
\vspace{0.3cm}
\hspace{3.2cm}
\epsfxsize=6cm
\epsfysize=5cm
\epsffile{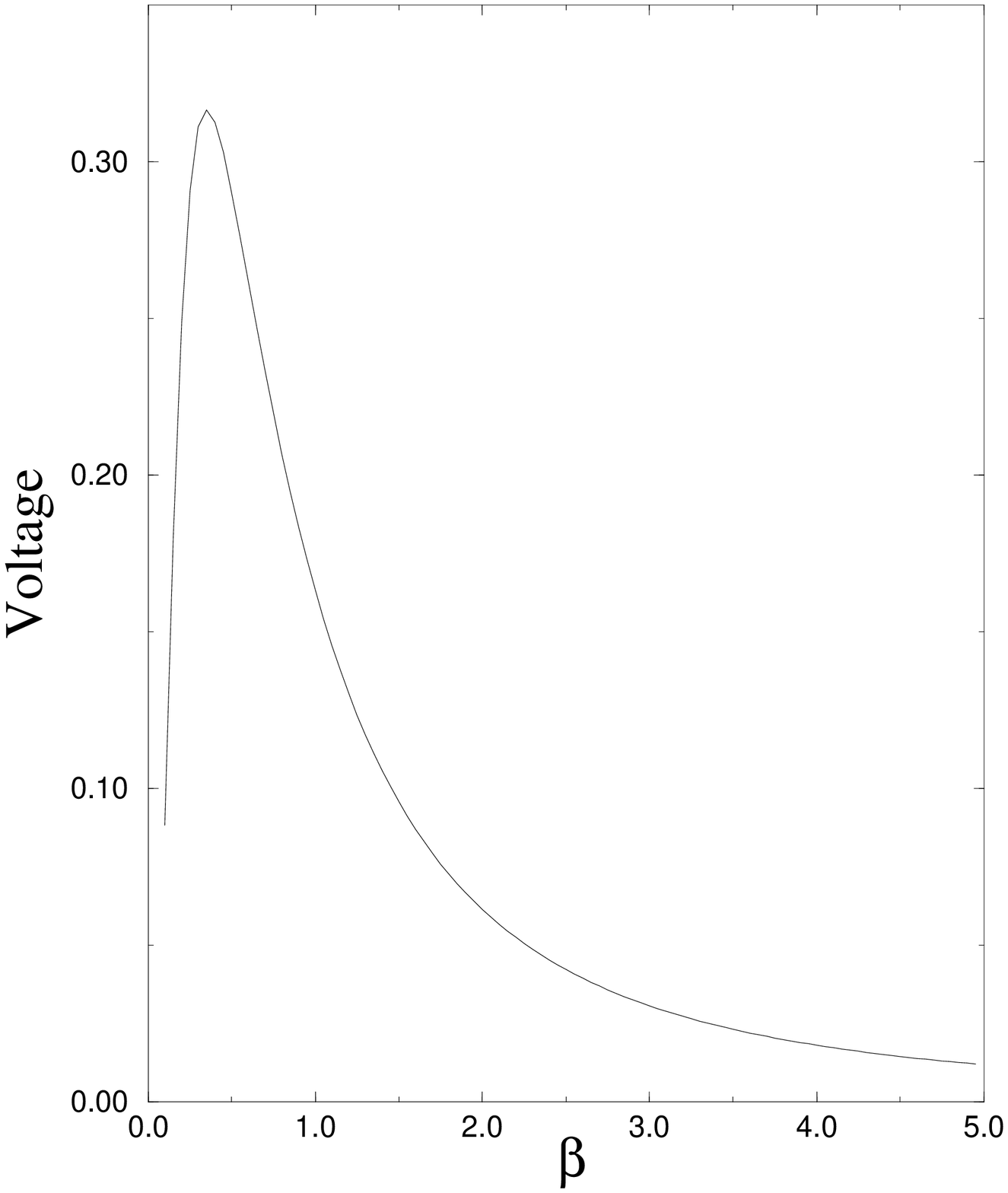}
\vspace{0.3cm}
\refstepcounter{figure}
\label{fig5}
\end{figure}
{\small \setlength{\baselineskip}{10pt} FIG.\ \ref{fig5} 
 The dependence  of the normalised voltage $\widetilde{V_{T}}$ on
inverse temperature $\beta $ at $\varphi =\pi /2$ (the parameters are $
\Delta /\epsilon _{Th}=10,L_1/L=0.5$).}

\end{document}